\documentclass[useAMS,usenatbib]{mn2e}
\usepackage{epsfig}
\usepackage{amsmath}
\usepackage[authoryear]{natbib}
\usepackage{longtable}
\usepackage{graphicx}
\usepackage{amssymb}
\def\mpc{h^{-1} {\rm{Mpc}}}

\title[II. The nature of assembly bias]
  {The nature of assembly bias - II. 
Halo spin}
\author[Lacerna $\&$ Padilla]
  {Ivan Lacerna$^{1, 2}$ and Nelson Padilla$^{1, 3}$\\
    $^1$Departamento de Astronom\'ia y Astrof\'isica, Pontificia Universidad Cat\'olica de Chile, 
V. Mackenna 4860, Santiago, Chile\\
    $^2$Instituto de Astronom\'ia, Universidad Nacional Aut\'onoma de M\'exico, A. P. 70-264, 04510,
    M\'exico, D.F., M\'exico \\
    $^3$Centro de Astro-Ingenier\'ia, Pontificia Universidad Cat\'olica de Chile, 
V. Mackenna 4860, Santiago, Chile\\    
}

\date{}

\pagerange{\pageref{firstpage}--\pageref{lastpage}} \pubyear{000}

\begin{document}

\label{firstpage}

\maketitle

\begin{abstract}
We study an assembly-type bias 
parametrized by the dimensionless 
spin parameter 
that affects massive
structures. In numerical simulations 
higher spin haloes are more strongly
clustered than lower spin haloes of equal mass.
We detect a difference of over a 30 per cent in the clustering strength for dark matter haloes of 10$^{13}$ - 10$^{14}$ $h^{-1}$ M$_{\odot}$, which is similar
to the result of Bett et al.
We explore whether the dependence of clustering strength on halo spin
is removed if we apply the redefinition of overdensity peak
height proposed by Lacerna \& Padilla (Paper I) obtained using assembly ages.
We find that this is not the case due to
two reasons. 
Firstly, only a few objects of low-virial mass are 
moved into the mass
range where the spin introduces an assembly bias after using
this redefinition.
Secondly, this formalism does not alter the mass 
of massive objects. In other words, the sample of haloes
with 
redefined mass $M$ in the high-mass regime is practically
the same as before the redefinition of peak height, 
and thus
the clustering behaviour is the same. 
We then repeat the process of finding the redefined peak height of Paper I but using the spin.
In this case, the new masses
show no spin-related assembly bias but they
introduce a previously
absent 
assembly bias with respect to relative age. 
From this result, we conclude that the assembly-type bias with respect to the halo spin 
has a different origin than with respect to assembly age.
The former 
may be due to the material
from filaments, which is accreted by massive haloes, that is enhanced in high-density environments, thus causing more extreme spin values
without significantly changing the formation age of the halo.
In addition, the 
estimates of 
the mass of collapsed structures in
numerical simulations could be lower than the 
true mass,
even in cluster-size haloes. 
High-mass objects 
may correspond, in some cases, to a different  
peak height than that suggested by their virial mass, 
providing a possible
explanation for the assembly bias with respect to spin.

\end{abstract}
\begin{keywords}
cosmology: theory - dark matter - large-scale structure of Universe. 
\end{keywords}
\section{Introduction}

The new generation of numerical simulations of high resolution have shown 
that the large-scale clustering of haloes of a given mass varies significantly with their assembly history
(Gao $\&$ White 2007). 
This effect, 
which is not expected from the  
extended Press$-$Schechter theory
(Bond et al. 1991), 
was termed `assembly bias', 
and is found when measuring the amplitude
of clustering as a function of halo properties such as
formation time and halo spin at fixed halo mass.

In the first paper of this series, we 
presented a new approach to estimate the overdensity peak height
with the aim to understand the assembly bias effect
(Lacerna \& Padilla 2011, hereafter Paper I). 
The method consisted in
redefining the 
overdensity that characterizes each object
using the information of its virial mass and 
\mbox{relative} age. 
This new definition is proposed as a better alternative than the virial mass as a proxy for equivalent peak height.

In this letter we will investigate the prevalence of the
assembly bias with the proposed redefinition
of an overdensity peak height of Paper I using the dimensionless spin parameter.
Works by Bett et al. (2007, hereafter B07) 
and
Faltenbacher \& White (2010) have detected  an assembly-type
bias in massive dark matter (sub)haloes using this property. They 
found that
higher spin objects are more strongly clustered than lower
spin objects of equal mass.

The outline of this work is as follows.
Section \ref{sec_data} describes the data,
and the dimensionless spin parameter
is defined in Section \ref{sub_spin_parameter}.
The detection of the assembly bias in
our simulation using halo spin
is shown in Section \ref{section_two-point_spin},
and Section \ref{sec_prm} shows the results using the overdensity
peak height proxy.
The discussion and conclusions are presented in Section \ref{reasons}.

\section{Data}
\label{sec_data}

The numerical simulation 
we use in this letter 
is called \verb|STAND|, 
and consists of a periodic box of 150 $h^{-1}$ Mpc 
on a side.
It contains $640^3$ dark matter particles with a mass resolution of $\sim$$10^9$ $h^{-1}$ $M_{\odot}$.
The cosmology used in this simulation
is $\Omega_m$ = 0.28 (with a baryon fraction of 0.164), $\Omega_\Lambda$ = 0.72, $\sigma_8$ = 0.81, $h$ = 0.7,
and spectral index $n_s$ = 0.96.
The chosen value of $\sigma_8$ is in good agreement with recent results of WMAP (Komatsu et al. 2011).
The simulation is run from redshift $z$ = 73.5 using the initial conditions of \verb|GRAFIC2| (Bertschinger 2001)
and the public version of the \verb|GADGET-2| code (Springel 2005).

In 99 snapshots of the simulation, dark matter haloes are 
identified as 
structures
that contain at least 10 particles using a
friends-of-friends (FOF) algorithm (Davis et al. 1985). 
Another algorithm (SUBFIND; Springel et al. 2001) is applied
to these groups in order to find substructures with at least 10 particles.

The catalogues of haloes and subhaloes are used to
construct merger histories, over which 
the SAG2 model by Lagos, Cora, $\&$ Padilla (2008; 
see also Lagos, Padilla, \& Cora 2009)
is run to produce a 
galaxy population. 
In this letter, we only use the central galaxy of the 
most massive subhalo
within a FOF halo.

\section{Dimensionless spin parameter}
\label{sub_spin_parameter}

The dimensionless spin parameter $\lambda$ is typically defined as (Peebles 1971)
\begin{equation}
\lambda = \frac{J |E|^{1/2}}{G M_h^{5/2}} ,
\label{eq_spin_Peebles}
\end{equation}
\\
where $J$ is the magnitude of the angular momentum vector, $E$ is the total energy, $G$ is
the gravitational constant, and $M_h$ is the mass of the FOF halo defined by the number of dark matter particles,
$N_p$.
%
%
The angular momentum is
\begin{equation}
\mathbf{J} = \frac{M_h}{N_p} \sum_{i=1}^{N_p} \mathbf{r}_i \times \mathbf{v}_i
\end{equation}
where $\mathbf{r}_i$ and  $\mathbf{v}_i$ are the position and velocity vectors of particle $i$ relative to the halo centre of mass, respectively.
Furthermore, the total energy of the system is given by $E = T + U$, where
$T$ corresponds to the kinetic energy and $U$ 
to the potential energy.

We select haloes with number of particles $N_p \ge$ 300 because they have reliable estimates of this parameter (B07).
The number of selected objects using this condition is 31,449 haloes, which corresponds to four per cent of the total sample. 
As noted by B07, 
there is an important scatter at higher spins 
even for $N_p \geq 300$
due to
small objects with high velocity dispersions, caused by the proximity of more-massive haloes.
This velocity contamination increases the kinetic energy of small haloes, $T$,
resulting in larger values of spin.
To overcome this problem, they suggested the quasi-equilibrium criterion to remove
anomalous spin haloes. This is a cut of the instantaneous virial ratio of halo energies
$2T/U + 1$.
Typically, a virialized object has a value around zero, and a gravitationally bound object
has value $\geq -1$. We 
select the haloes according to

\begin{equation}
-Q \leq \frac{2T}{U} + 1 \leq Q,
\label{eq_TU}
\end{equation}
\\
where the limit value is $Q = 0.5$ 
(B07).
This allows us to remove objects with anomalous spins.
The final sample contains 29,633 haloes.

Although the spin $\lambda$ has 
little dependence 
on mass, if any, we define the relative
spin parameter, in analogy to the relative age parameter $\delta_t$ of Paper I, as

\begin{equation}
\delta_{\lambda_{i}} = \frac{\lambda_{i}- \big<\lambda(M)\big>}{\sigma_\lambda(M)} ,
\label{eq_delta_spin} 
\end{equation}
\\
where, for the $i$ th object, $\lambda_{i}$ is its dimensionless spin, 
$\big<\lambda(M)\big>$ is the
median spin as a function of FOF host halo mass $M$,
and $\sigma_\lambda(M)$ is the dispersion around the median.
Then, positive values of $\delta_\lambda$ correspond to high spin objects, whereas negative values of $\delta_\lambda$
are related to low spin objects. 


\section{Assembly bias using halo spin}

\label{section_two-point_spin}

In this section, we will use the 
two-point correlation function
to test whether 
dark matter (DM) haloes from the STAND simulation show an
assembly-type bias with the halo spin.

\begin{figure}
\begin{center}
\leavevmode \epsfysize=8.5cm \epsfbox{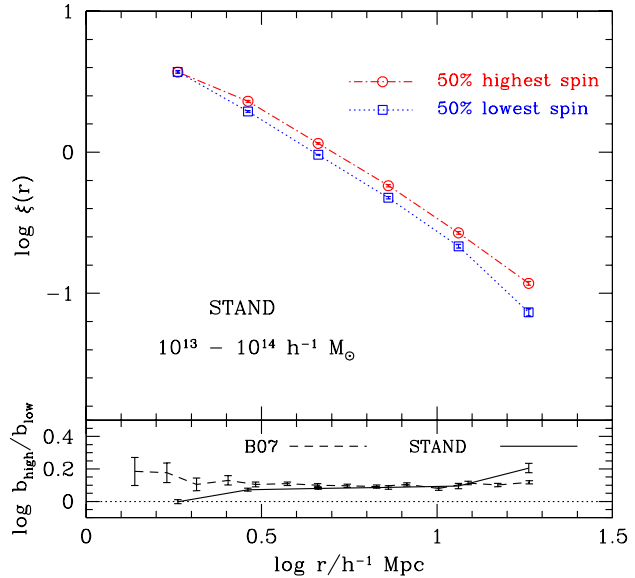} 
\caption [Two-point cross-correlation function of haloes of equal mass but different spin]
{Main panel (top): two-point cross-correlation function for haloes 
in the STAND simulation.
The result for the 50 per cent highest spin haloes is represented as dot-dashed red lines, whereas that for the 50 per cent lowest spin haloes
appears as dotted blue lines.
Error bars were calculated using the jackknife method.
Lower panel (bottom): ratio between the bias of high and low spin objects
in the STAND simulation (solid lines) and in B07 (dashed lines).
At large scales, $r > 3$ $\mpc$, both simulations show a higher clustering for
high spin haloes with respect to low spin ones with a similar signal, which is over 30 per cent.}
\label{xi_bin3_spin_STAND}
\end{center}
\end{figure}

Figure \ref{xi_bin3_spin_STAND} shows the two-point cross-correlation function of DM haloes
of equal mass but different relative spin parameter given by equation (\ref{eq_delta_spin}).
The centres to perform $\xi(r)$ are haloes
selected in the mass range 10$^{13}$ - 10$^{14}$ $h^{-1}$ M$_{\odot}$, and the tracers are all the haloes in the final
sample (see Section \ref{sub_spin_parameter}) in order to obtain
a better signal-to-noise.
The 50 per cent 
high spin haloes are more strongly clustered at large scales than
the 50 per cent 
low spin haloes, thus showing an assembly-type bias.
The lower panel shows the 
ratio between the bias of high and low spin objects
in the STAND simulation (solid line). 
This ratio is calculated as 
\begin{equation} 
\frac{b_{H,high_\lambda}}{b_{H,low_\lambda}}= \frac{\xi_{HH',high_\lambda}}{\xi_{HH',low_\lambda}},
\end{equation}
where the subscript $H$ refers to 
centres
and $H'$ 
to tracers.
We compare to 
the autocorrelation function results of B07 shown as a dashed line in the lower panel.
In this case, the ratio between the bias of 50 per cent high and 50 per cent low spin objects
is calculated as
\begin{equation} 
\frac{b_{H,high_\lambda}}{b_{H,low_\lambda}}= \sqrt{\frac{\xi_{HH,high_\lambda}}{\xi_{HH,low_\lambda}}}.
\end{equation}
As can be seen from the lower panel, at scales $r > 3$ $\mpc$, both simulations show a higher clustering for
high spin haloes with respect to the low spin ones 
(over a 30 per cent difference) with a comparable 
signal.

\section{Overdensity peak height proxy }
\label{sec_prm}

In Paper I, we presented an approach to trace the assembly bias effect.
This consisted in measuring the total mass inside spheres of 
different radii around semi-analytic galaxies, which in some cases 
returned larger values 
than the virial mass of the host dark matter halo, using two free parameters
to introduce a dependence of this radius on mass and age.
This model defined a new overdensity peak height for which the 
large-scale clustering of objects of a given mass did not depend on the age.  
The parametrization 
helps to characterize the environment 
and to understand what is behind the assembly bias
(see Paper I for more details).

\subsection{Parametrization using the relative age}
\label{sub_prm_age}

We will first use the same approach 
of Paper I 
in order to find the best-fitting parameters
so that $\xi(r)$ does not show a dependence of the large-scale clustering on age. We will then use 
the parametrization to study whether the dependence on spin is still present. 

For this purpose, we match up the position of central semi-analytic galaxies 
with the position of the selected haloes in order to estimate the relative age of these haloes
using 
\begin{equation}
\delta_{t_{i}} = \frac{t_{i}- \big<t(M)\big>}{\sigma_t(M)} ,
\label{eq_delta} 
\end{equation}
\\
where, for the $i$ th galaxy, $t_{i}$ is its mass-weighted stellar age, 
$\big<t(M)\big>$ is the
median stellar age as a function of host halo mass $M$, 
and $\sigma_t(M)$ the dispersion around the median in units of time.
The procedure is successful in the 99 per cent of the cases,
so that the final 
sample  
contains 29,376 haloes for which
the results shown in Figure \ref{xi_bin3_spin_STAND}
are the same.
In the other 257 objects, 
which include some
massive haloes of $10^{14.5}$ $h^{-1} M_{\odot}$, 
the position of the central galaxy is very far away
from the central postion of the halo. We use the condition $d \leq$ 250 $h^{-1}$ kpc, where $d$ is the distance
between the position of the central galaxy and centre of mass of the DM halo. We do not 
assign the 
age of the central galaxy 
to its host halo at distances beyond this limit.
Again, these `failures' correspond to just one per cent of the haloes.
The resulting stellar age as a function 
of halo mass is qualitatively similar to that shown in 
Paper I. 
On average, more massive objects have older stellar ages.

\begin{figure}
\begin{center}
\leavevmode \epsfysize=8.5cm \epsfbox{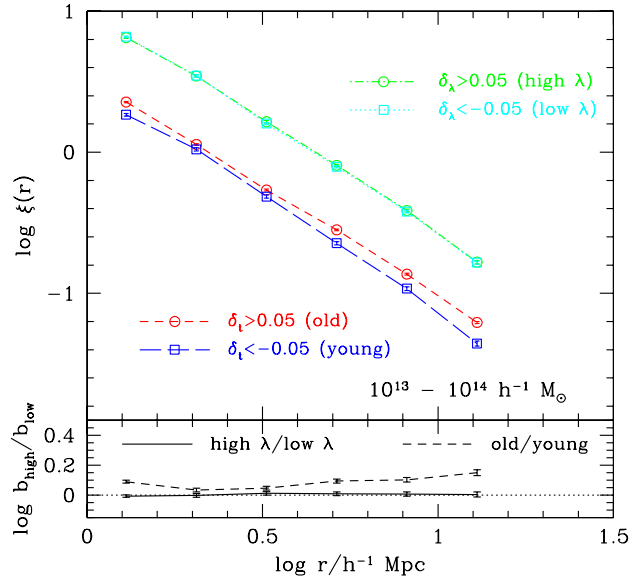} 
\caption 
{Two-point cross-correlation function for haloes of equal redefined mass
but different spin
and different age 
from the STAND simulation.
The best-fitting parameters are $a=-0.36$ and $b=-1.52$,
where the reduced $\chi^2$
statistics is performed using the spin 
(Section \ref{sub_prm_spin}).
Error bars are calculated using the jackknife method.
The high-spin population is represented as a dot-dashed green line, whereas the low-spin population is represented as a dotted cyan line.
The old population is represented as a short-dashed red line, whereas the young one appears as a long-dashed blue line. For clarity, these two lines are moved -0.5 dex in the vertical direction.
Lower panel: ratio between the bias of high-spin and low-spin objects (solid line)
and between the bias of old and young objects (dashed line).
The dotted line represents the ratio equal to unity.
}
\label{S15_bin3}
\end{center}
\end{figure}

We redefine the overdensity peak height within spheres around
haloes of mass $M_h$, for which the radius of a sphere (in $\mpc$ units) 
is parametrized as 
\begin{equation}
r = a \textrm{ $\delta_t$} + b\textrm{ log}\left(\frac{M_h}{M_{nl}}\right) \textrm{ ,} 
\label{eq_r_prm}
\end{equation}
\\
where $M_{nl}$ is the non-linear mass defined by Seljak \& Warren 
(2004),  
log($M_{nl}$/$h^{-1}$ M$_{\odot}$) = 13.24 
for the choice of cosmological parameters in the STAND simulation.
The free parameters are $a$ and $b$. 
The new peak height proxy will be the mass $M$ enclosed within this radius.
It is assumed that if $r$ is smaller than the virial radius 
or if $M$ is smaller than the halo mass, 
then $M = M_{h}$.
The $\chi^2$ statistics to find the best-fitting set of free parameters is defined as
\begin{equation}
\chi^2_{\xi(r)} = \frac{1}{N} \sum_{i}^N \left(\frac{1}{n_{dof}} \sum_r \frac{ \big[\xi_{high}(r) - \xi_{low}(r) \big]^2}{\sigma^2}\right)_i 
\textrm{ ,}
\label{eq_chixi}
\end{equation}
where $\xi_{high}(r)$ and $\xi_{low}(r)$ is the result for old and young haloes, respectively.
The error is $\sigma^2 = \sigma^2_{\xi_{high}} + \sigma^2_{\xi_{low}}$,
where the first term is
the jackknife error for $\xi_{high}(r)$ and the second for $\xi_{low}(r)$.
The symbol $n_{dof}$ denotes the number of degrees of freedom ($n_{dof} = 6$).
This statistics is calculated within  1 $\leq r/h^{-1}$ Mpc $\leq 13$
over three mass bins (the first, second and third terciles of the mass distribution),
i.e. $ N = 3$.

The best-fitting values are $a=0.08$ and $b=-0.37$.
As in Paper I, the assembly
bias with respect to age is not present at large scales
after performing this procedure.
However, 
the differences in the clustering strength of
haloes of equal high mass but different spin
will still be present
because just one halo of low virial mass is considered in this high-mass range
after using the parametrization. Also, as mentioned in Paper I,
this formalism does not alter the mass of objects with high mass.
Therefore, the sample of haloes with 
masses 
between $10^{13}$ and $10^{14}$ $h^{-1}
M_{\odot}$ is practically the same, thus showing the same clustering behaviour as before.

\subsection{Parametrization using spin}
\label{sub_prm_spin}


To trace the assembly bias with 
the dimensionless spin parameter, we use  
the reduced $\chi^2$ statistics 
defined 
in eq. (\ref{eq_chixi}),
%
but in this case 
$\xi_{high}(r)$ and $\xi_{low}(r)$ 
correspond to
the cross-correlation function 
for high-spin and low-spin haloes, respectively.
This statistics is calculated within 1 $\leq r/h^{-1}$ Mpc $\leq 13$
over two mass bins, $10^{12} - 10^{13}$ $h^{-1}
M_{\odot}$ and $10^{13} - 10^{14}$ $h^{-1} M_{\odot}$, i.e. $N = 2$.

\begin{figure}
\begin{center}
\leavevmode \epsfysize=8.5cm \epsfbox{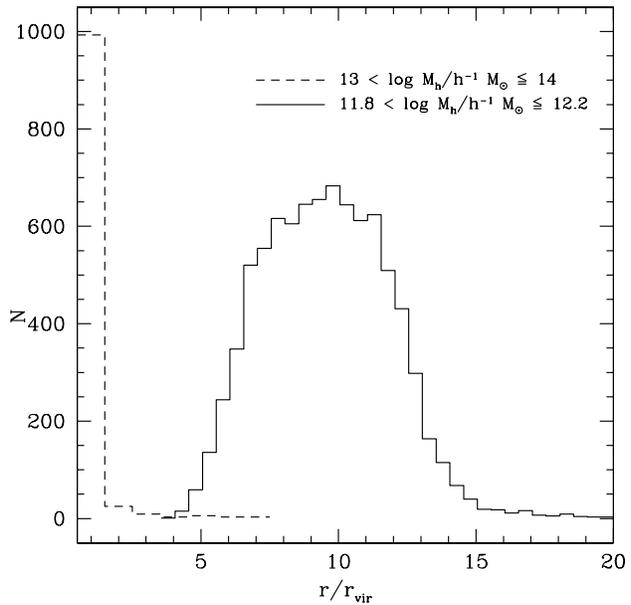} 
\caption 
{Distribution of radius $r$ 
of eq. (\ref{eq_r_prm})
in units of viral radius $r_{vir}$ for two ranges of
halo mass $M_h$ from the STAND simulation 
for the parametrization using spin
(see Section \ref{sub_prm_spin}).
Haloes of intermediate mass (solid line) have mean radius
of 10$r_{vir}$ , which is larger compared to 1 - 4$r_{vir}$ described in Paper I. 
Also, a few very massive objects (dashed line) have $r > r_{vir}$,
in contrast to the redefinition of the overdensity peak height 
that explains the assembly bias using the age.
}
\label{ratios_S15}
\end{center}
\end{figure}

The best-fitting
values are $a=-0.36$ and $b=-1.52$. 
After applying the redefinition of mass
there is no difference in the clustering 
amplitude for massive haloes of high and low spin at large scales,
as can be seen 
in Figure \ref{S15_bin3}.
In contrast, 
it can be seen in the same figure 
an artificial assembly bias with respect to age using this set of parameters. Note that the 
assembly bias for objects of equal mass but different
age 
affects low-mass haloes 
(Gao, Springel \& White 2005).
Therefore, the formalism applied to the assembly-type bias with the dimensionless spin parameter $\lambda$ is not related with the successful overdensity peak height
defined in Paper I. For example,
Figure \ref{ratios_S15} shows that the mean radius for haloes of intermediate mass
is 10$r_{vir}$ (solid line), which is larger compared to 1 - 4$r_{vir}$ described in Paper I. 
Furthermore, a few very massive objects changed their masses
(dashed line),
an aspect that was not found in the redefinition of the overdensity peak height 
that explains the assembly bias using the age.

\section{Discussion and Conclusions}
\label{reasons}

\begin{figure}
\begin{center}
\leavevmode \epsfysize=8.5cm \epsfbox{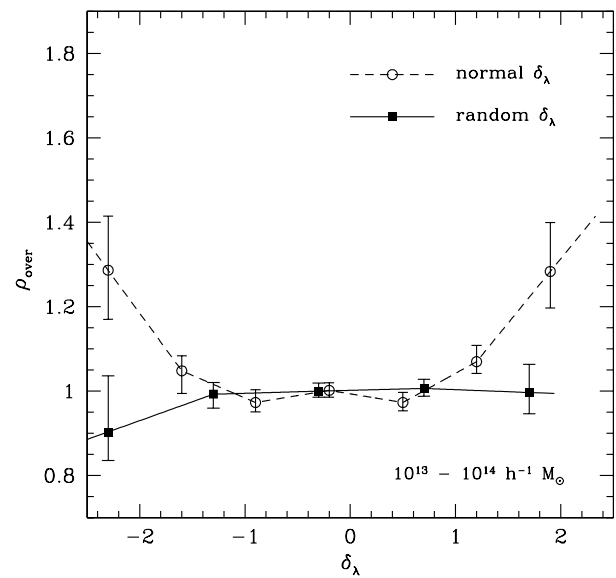} %
\caption {Median overdensity as a function of relative spin parameter, $\delta_{\lambda}$,
for 
massive haloes in the STAND simulation.
Open circles (dashed line) correspond to normal $\delta_{\lambda}$ values.
Filled squares (solid line) correspond to randomly swapped $\delta_{\lambda}$ values
(see details in Section \ref{reasons}).
The error bars correspond
to the 
error of the median in both cases.
}
\label{overdens_spin_massive}
\end{center}
\end{figure}

Perhaps the assembly-type bias found in massive objects of equal mass but different
spin values is related with the massive neighbours responsible 
for the truncation of the growth
of smaller objects reported in Paper I. 
High mass haloes reach virial masses according to theoretical expectations, but those in high density environment accrete an
important fraction of material from filaments (Hahn et al. 2009). This highly
collimated material 
could cause changes in the angular momentum of these haloes
affecting 
their spin. On the other hand, 
low-mass objects of equal mass do not show important differences in clustering when they are split in samples of high and low spin $\lambda$ 
showing a lack of dependence with the environment (B07). 
However, we have already mentioned that small haloes in the vicinities of
massive neighbours are discarded in this kind of analysis since they suffer velocity contamination that increases their kinetic energy, thus increasing their spin values.
Probably, massive haloes fed 
by material in the region of a truncated smaller halo suffer a similar contamination that alters their spins. 

The environmental dependence of massive (13 $\le$ log($M_{h}/h^{-1}$ M$_{\odot}$) $\le$ 14) haloes is shown in Figure \ref{overdens_spin_massive}.
This dependence is estimated as the 
median overdensity ($\rho_{over}$)
as a function of relative spin parameter, $\delta_{\lambda}$. 
The density is calculated using all the 
available haloes in the simulation.
We then estimate the overdensity with respect to the median. 
As can be seen, 
a smooth dependence with environment is found for low-spin and high-spin objects of equal mass (dashed line).
It is worth to mention that extreme values of spin are represented by
higher or lower values of $\delta_{\lambda}$. Therefore, it is plausible
that the assembly-type bias found by using the spin parameter is due to a density enhancement
that affects the spin values of massive objects. 

\begin{figure}
\begin{center}
\leavevmode \epsfysize=8.5cm \epsfbox{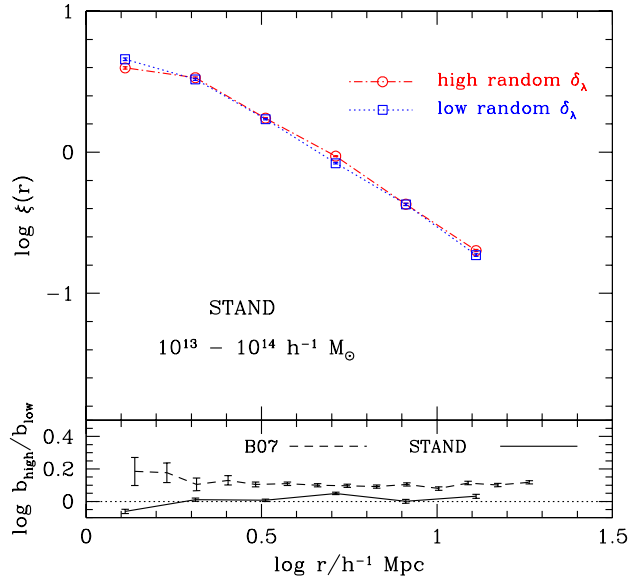}  
\caption
{
$\xi(r)$ for massive haloes
from the STAND simulation.
Relative spin parameter $\delta_{\lambda}$ is randomly 
swapped among haloes in mass ranges of 0.1 dex.
The result for 50 per cent highest spin haloes is represented as dot-dashed red lines, whereas that for 50 per cent lowest spin haloes
appears as dotted blue lines.
Error bars are calculated using the jackknife method.
Lower box: ratio between the bias of high and low spin objects
in the STAND simulation (solid line) and in B07 (dashed line).
The average difference in clustering amplitude
is smaller than a 10 per cent when halo spins of the STAND simulation are
randomly distributed.
}
\label{xi_bin3_randomspin_STAND}
\end{center}
\end{figure}

In order to 
remove this dependence with environment, the relative spin parameter $\delta_{\lambda}$ is randomly 
swapped among 
haloes in mass ranges of 0.1 dex
(see the resulting solid line in Figure \ref{overdens_spin_massive}), 
without altering their positions.
After performing this procedure, the 
halo clustering dependence on spin is practically
absent, as can
be seen in Figure \ref{xi_bin3_randomspin_STAND}.
The average difference in clustering strength between the fifty per cent highest spin haloes
and
the fifty per cent lowest spin haloes
is smaller than a 10 per cent,
showing that the density enhancement can affect 
the properties of massive objects such as the spin.

Therefore, in the context of the peak formalism,  the 
large amount of material from 
filaments from the cosmic web 
that a massive halo accretes in 
high density regions might also affect
its final mass, thus 
introducing a spurious 
apparent change in the bias.
Furthermore, tools used to identify objects in numerical simulations, such as FOF or SUBFIND, often define structures located very close
to a massive object as a separated identity. 
Anderhalden \& Diemand (2011) developed a formalism
to identify lost particles, where many small structures actually
belong to a neighbour massive object. As a consequence,
the mass estimation of structures in numerical
simulations, even in cluster-size haloes, could be
lower than the actual mass.
It is remarkable that the age of massive objects
is not strongly affected by this 
discrepancy in mass;
however
this 
phenomenon
is important in properties that are directly related
with groups of DM particles in high density environments
such as the spin,
which can produce an assembly bias effect.

\bigskip


We would like to thank Darren Croton, Gaspar Galaz, Leopoldo Infante and the anonymous referee for comments and discussions.
We acknowledge support from FONDAP ``Centro de Astrof\'\i sica" $15010003$, BASAL-CATA,
Fondecyt grant No. 1110328, CONICYT, and MECESUP.
IL acknowledges support from DGAPA-UNAM.
The calculations for this work were performed using the Geryon cluster at the Centre for Astro-Engineering at UC.

\label{lastpage}

\end{document}